\documentclass[12pt]{article}
\usepackage{amssymb}
\usepackage{amsfonts}
\usepackage{amsmath}
\usepackage{amsmath,amsthm}
\usepackage{graphicx,epsfig, color }

\oddsidemargin 10mm \numberwithin{equation}{section}
\def\be{\begin{equation}}
\def\ee{\end{equation}}
\begin{document}
\begin{center} {{\bf {Dynamical system approach to scalar-vector-tensor cosmology}}\\
 \vskip 0.10 cm
 {{H. Ghaffarnejad\footnote{E-mail address: hghafarnejad@seman.ac.ir}} and E. Yaraie\footnote{E-mail address:
 eyaraie@gmail.com}}}
 \vskip 0.1 cm
 \textit{Faculty of Physics, Semnan
University, Semnan, Zip Code: 35131-19111, IRAN}
\end{center}
\begin{abstract}
 Using scalar-vector-tensor Brans
Dicke (VBD) gravity [3] in presence of self interaction BD
potential $V(\phi)$ and perfect fluid matter field action we solve
corresponding field equations via dynamical system approach for
flat  Friedmann Robertson Walker metric (FRW). We obtained 3 type
critical points for $\Lambda CDM$ vacuum de Sitter era  where
stability of our solutions are depended to choose particular
values of  BD parameter $\omega.$ One of these fixed points is
supported by a constant potential which is stable for $\omega<0$
and behaves as saddle (quasi stable) for $\omega\geq0.$ Two other
ones are supported by a linear potential $V(\phi)\sim\phi$ which
one of them is stable for $\omega=0.27647.$ For a fixed value of
$\omega$ there is at least 2 out of 3 critical points reaching to
a unique critical point. Namely for $\omega=-0.16856(-0.56038)$
the second (third)  critical point become unique with the first
critical point. In dust and radiation eras we obtained 1 critical
point which never become unique fixed point. In the latter case
coordinates of fixed points are also depended to $\omega.$ To
determine stability of our solutions we calculate eigenvalues of
Jacobi matrix of 4D phase space dynamical field equations  for de
Sitter, dust and radiation eras. We should be point also
potentials which support dust and radiation eras must be similar
to $V(\phi)\sim\phi^{-\frac{1}{2}}$ and $V(\phi)\sim\phi^{-1}$
respectively. In short our study predicts that radiation and dust
eras of our VBD-FRW cosmology transmit to stable de Sitter state
via non-constant potential (effective variable cosmological
parameter) by choosing $\omega=0.27647$.
\end{abstract}
\section{Introduction}
The physical nature itself is complex system really and is
described by nonlinear chaotic dynamics [4,5,6]. A chaotic
dynamics in its continues (discrete) form is described by
nonlinear (iteration maps) differential equations. It leads
usually to its possible stable points called as attractors (see
arrow diagrams in figure 1). Usually dynamical systems
 described by nonlinear differential equations have not regular
analytic solutions. This restrict us to choose geometrical
approach to solve them. The latter method gives us properties of
the solutions without the solutions themselves. Properties of the
solutions are called as attractors (sink and/or stable) and
saddles (quasi-stable).
  The phase space variables of dynamical system at the
classical mechanics are well known as canonical coordinates and
corresponding momenta but not in the cosmological context. In the
latter case dynamical variables are more and there are several
degrees of freedom to choose them. If we choose unsuitable choices
so can not obtain physically applicable solutions according to the
experimental context. This restrict us to regard two important
statements about the geometrical variables as must be
dimensionless and bounded. The latter two properties make as
finite the phase space which means all of the critical points
become
 visible.\\
Choosing some suitable dimensionless geometrical variables one can
reduces a $`n`$ order nonlinear differential equation of a
dynamical system to number of $`n`$ to first order differential
equations as
\begin{equation}
 \overrightarrow{\dot{x}}=\frac{d\overrightarrow{x}}{dt}=f(\overrightarrow{x},t)
 \end{equation}
where $\overrightarrow{x}=\{x^i;i=1,2,3,....n\}\in E\subseteq R^n$
is state of $`n`-$dimensional phase space $E\subseteq R^n.$ The
equation (1.1) is called as autonomous if
\begin{equation} \frac{\partial f}{\partial t}=0
   \end{equation}
 and non-autonomous if \begin{equation} \frac{\partial f}{\partial t}\neq0.\end{equation}
Solutions of the equation $\overrightarrow{\dot{x}}=0$ gives us
critical points $P_c$ of the dynamical system. One can obtain
eigenvalues $\lambda_i$ of each critical point by calculating
Jacobi matrix of the vector function $f(\overrightarrow{x},t)$
defined by
\begin{equation}
[\mathcal{J}]_{ij}=J_{ij}=\bigg(\frac{\partial \dot{x}^i}{\partial
x^j}\bigg)_{x=x_{critical}}\end{equation} and solving its secular
equation  as\be\det(J_{ij}-\lambda\delta_{ij})=0.\ee This leads to
an algebraic equation which its solutions give us eigenvalues of a
critical point of the dynamical system. Characters of obtained
critical points are depended to numerical value of the
corresponding eigenvalues. For instance, the critical point is
called as unstable (repeller and/or source) if the corresponding
eigenvalues take positive real value numerically. If at least one
of all real eigenvalues takes negative real value numerically then
the critical point is called saddle. The critical points is called
as stable (attractor and/or sink), if all of the eigenvalues take
negative real value. If eigenvalues take complex numbers with
positive (negative) real value then the critical point is called
as spiral unstable
 (stable). For zero eigenvalues the system  become degenerated and so we can not tell
 about stability and/or instability of the dynamical system under
 consideration (see table 1).\\
 In context of cosmological models the dynamical system approach is used  to obtain $\Lambda$CDM phase
 by more authors [7-17]:  Zhou et al are used $f(G)$ gravity  to study flat FRW
 cosmology  in [7]
  where $G=R^2-4R^{\mu\nu}R_{\mu\nu}+R^{\mu\nu\lambda\eta}R_{\mu\nu\lambda\eta}$
  is Gauss-Bonnet topological invariant.
   They are obtained two kinds of stable
   accelerated solutions called as de Sitter and phantom-like of dark energy regime.
   Azizi and Yaraie are used non-minimally matter coupled $f(R)$  gravity
   to study flat FRW
   cosmology in [8]. They are obtained vacuum de Sitter era of the Universe which can be mimic
   the late-time acceleration of the cosmic evolution. Hrycyna and
   Szydtowski are studied BD-FRW cosmology in presence of a quadratic scalar potential
   in [9,10] containing stable de Sitter phase and are studied observational constraints of the model in ref. [11].
      Copeland et al analyzed the dynamics of a single scalar field in
FRW universes with spatial curvature in ref. [12] where an
attractor critical point is obtained to satisfy de Sitter and
power law expanding Universe.  Matos et al studied dynamical
approach of scalar-tensor cosmology in presence of a $\cosh$ type
of the potential plus a cosmological constant reducing to de
Sitter attractor [13]. Lopez and Ibarra studied attractor
properties of the chaotic inflation in [14]  by using a minimally
coupled scalar field in presence of a quadratic scalar potential.
Amendola was used quintessence (light scalar field) effects to
study cosmic acceleration via dynamical system approach in
presence of dynamical cosmological parameter and an exponential
potential [15]. He obtained a multi-pole spectrum effect of the
microwave background at large angles where the acoustic peaks are
shifted and their amplitude is changed.  Fay et al are
   obtained particular class of $f(R)$ modified gravity theories which can be
   mimic $\Lambda CDM$ cosmology in ref. [16]. Nozari and Kiani are
   studied (1+4) dimensional bran-world cosmology containing a Gauss-Bonnet term at
   bulk action and obtained stable de Sitter phase state  [17].
   In short, we know that $\Lambda$CDM phase of accelerating Universe is supported via ansatz of unknown cosmological constant
   $\Lambda$ in general theory of relativity and dark matter inflaton scalar field in more scalar tensor gravity theories.
   Really, origin of the non-baryonic dark matter proposal is not known and there are more candidate
     for it [18,19]. Diversity of dark matter particles candidate
     encouraged more authors to present alternative models as
     scalar-vector-tensor gravity theories (TeVeS) without non-baryonic dark matter
     which one can use instead of the general theory of relativity itself and/or usual scalar tensor
     models. In these models dynamical vector fields are four velocity of preferred
reference frames satisfying general covariance condition. These
vector fields support acceleration of the expanding Universe,
galaxy rotation curves
     and corrections on gravitational acceleration law in solar system, astrophysical and cosmological scales [20]
     instead of less-known non-baryonic dark matter (see also refs. [21,22] for their
     experimental constraints). If dynamical vector fields to be have unit-time-like property then the
     scalar-vector-tensor gravities can be also support metric
signature
      transition dynamics from Euclidean (+,+,+,+) to Lorentzian (-,+,+,+) signature (see  [3] and reference therein).
  In short, the model presented in
ref. [3] is generalized BD gravity [2] by transforming the
background metric $g_{\mu\nu}$ to $g_{\mu\nu}+2N_{\mu}N_{\nu}.$
 Flat FRW quantum cosmology
of the model and its metric signature transition property were
studied in refs. [23] and [24] respectively.\\
In this paper we use dynamical system approach of the gravity
model [3] in presence of self-interacting BD potential and
matter-radiation perfect fluid cosmic source and obtain $\Lambda$
CDM de Sitter stable phase of the accelerating flat
 FRW Universe. Originally, vector field stress tensor of our used model which support inflation of the cosmological Universe
 makes free of Jordan and/or Einstein frame of the used BD
 gravity.
 While Salcedo et al is shown in ref. [1] that the scalar
tensor Brans Dicke (SBD) gravity itself [2] in presence of
quadratic self-interaction potential their attractor de Sitter
solution is only valid for Jordan frame. Hence they claimed that
the BD gravity itself dose have not a $\Lambda CDM$ phase as an
universal attractor. Form the latter view our work can be
outstanding and so considerable to study with more details. As an
experimental result they obtained time variation of the Newton`s
gravitational coupling parameter as $|\dot{G}/G|<9\times10^{-13}$
$yr^{-1}$ for experimental values of Hubble constant
$H_0=7.24\times10^{-11}$ $yr^{-1}$ and BD parameter $\omega=40000$
while we obtained  its corrections coming from preferred reference
frame effects (dynamical vector fields corrections).
  Organization of the work is as
follows.\\ In section 2 we call the gravity model [3] and
calculate its dynamical field equations. In section 3 we obtain
Friedmann equations of the model. Next we make 4D cosmic dynamical
phase space to write corresponding dynamical equations. Then we
obtain critical points, matrix Jacobi and their eigenvalues  for
$\Lambda CDM$ vacuum de Sitter, dust and radiation  eras. Finally
we denote to concluding remark in section 4.
\section{The Model}
 Let us we start with the following scalar-vector-tensor-gravity action
[3].
 \be I_{total}=I_{BD}+I_{N}+I_{m}+I_{r}\ee which with assumption $\epsilon=0$ the term \be I_{BD}=\frac{1
 }{16\pi}\int dx^4\sqrt{g}\left\{\phi R-\frac{\omega}{
 \phi}
 g^{\mu\nu}\nabla_{\mu}
 \phi\nabla_{\nu}\phi+V(\phi) \right\}\ee   is BD scalar tensor
 gravity itself [2]. $V(\phi)$ is called as BD self interaction potential and \be I_N=\frac{1}{16\pi}\int
 dx^4\sqrt{g}\{\zeta(x^{\nu})(g^{\mu\nu}N_{\mu}N_{\nu}+1)+2\phi F_{\mu\nu}F^{\mu\nu}$$$$-\phi N_\mu
 N^{\nu}(2F^{\mu\lambda}\Omega_{\nu\lambda}+
 F^{\mu\lambda}F_{\nu\lambda}+\Omega^{\mu\lambda}\Omega_{\nu\lambda}
-
2R_{\mu}^{\nu}+\frac{2\omega}{\phi^2}\nabla_{\mu}\phi\nabla^{\nu}\phi)\},\ee
with   \be
F_{\mu\nu}=2(\nabla_{\mu}N_{\nu}-\nabla_{\nu}N_{\mu}),~~~~~~~\Omega_{\mu\nu}=2(\nabla_{\mu}N_{\nu}+\nabla_{\nu}N_{\mu})\ee
describes action of unit time like dynamical four velocity
$N_{\mu}(x^{\nu})$ of a preferred reference frame.  Up to
$\zeta(x^{\nu})$ term which is used as ansatz, the action (2.3) is
obtained by transforming metric field of the BD action (2.2) as
$g_{\mu\nu}\to g_{\mu\nu}+2N_{\mu}N_{\nu}.$  Details of
calculations  are given in ref. [3]. Matter and radiation
counterparts of a perfect fluid source is considered as
\begin{equation}I_m=\frac{1}{16\pi}\int
 dx^4\sqrt{g}L_m \end{equation} and \begin{equation}I_r=\frac{1}{16\pi}\int
 dx^4\sqrt{g}L_r\end{equation} where $L_m$ and $L_r$
  are the matter and radiation lagrangian densities respectively.  $T_{\mu\nu}$ is the matter-radiation stress
energy-momentum tensor and is given against the corresponding
Lagrangian density as follows.
\begin{equation}
T_{\mu\nu}=\frac{2}{\sqrt{g}}\frac{\delta(\sqrt{g}(L_m+L_r))}{\delta
g^{\mu\nu}}.
\end{equation}
We will assume $T_{\mu\nu}$ to be stress energy-momentum tensor of
a perfect fluid
 in what follows as \begin{equation} T_{\mu\nu}=(\rho+p)u_{\mu}u_{\nu}+pg_{\mu\nu}.
\end{equation}
Here $u^{\mu}$ is time-like four velocity of the fluid satisfying
$g^{\mu\nu}u_{\mu}u_{\nu}=-1.$ $\rho=\rho_m+\rho_r$ and
$p=p_m+p_r$ where $\rho_m(\rho_r)$ is matter (radiation)
counterpart energy density of the fluid and $p_m(p_r)$ is
corresponding isotropic hydrostatic pressure. The action (2.3)
shows that the vector field $N_{\mu}$ is coupled as non-minimally
with the BD scalar field $\phi.$ The action (2.1) is written in
units $c=\hbar=1$ with Lorentzian signature (-,+,+,+). The
undetermined Lagrange multiplier $\zeta(x^{\nu})$ controls that
$N_{\mu}$ to be an unit time-like vector field. $\phi$ describes
inverse of variable Newton`s gravitational coupling parameter and
its dimension is $(lenght)^{-2}$ in units $c=\hbar=1$. Absolute
value of determinant of the metric $g_{\mu\nu}$ is defined by $g$.
Present limits of dimensionless BD parameter $\omega$ based on
time-delay experiments [25,26,27,28] requires
$\omega\geq4\times10^{4}.$ General relativistic approach of the BD
gravity action (2.2) is obtained by setting $V(\phi)=0$ and
$\omega\to\infty$. Varying (2.1) with respect to $\zeta(x^{\nu}),$
$\phi,$ $N^{\mu}$ and $g^{\mu\nu}$ we obtain respectively   \be
g^{\mu\nu}N_{\mu}N_{\nu}=-1,\ee \be
\frac{2\omega\Box\phi}{\phi}-\frac{\omega
g^{\mu\nu}\partial_{\mu}\phi\partial_{\nu}\phi}{\phi^2}-\frac{4\omega
N^{\mu}N^{\nu}\partial_{\mu}(\sqrt{g}\partial_{\nu}\phi)}{\phi\sqrt{g}}-\frac{d
V(\phi)}{d \phi}-\frac{4\omega\partial_{\mu}(N^{\mu}N^{\nu})
\partial_{\nu}\phi}{\phi}$$$$-\frac{4\omega\Gamma^{\mu}_{\mu\alpha}N^{\alpha}N^{\nu}\partial_{\nu}\phi}{\phi}
-\frac{4\omega\Gamma^{\nu}_{\mu\lambda}N^{\mu}N^{\lambda}\partial_{\nu}\phi}{\phi}+\frac{2\omega
N^{\mu}N^{\nu}\partial_{\mu}\phi\partial_{\nu}\phi}{\phi^2}+R-2N^{\mu}N^{\nu}R_{\mu\nu}$$$$+2F_{\mu\nu}F^{\mu\nu}
-N_{\mu}N^{\nu}\{
2F^{\mu\lambda}\Omega_{\nu\lambda}+F^{\mu\lambda}F_{\nu\lambda}+
\Omega^{\mu\lambda}\Omega_{\nu\lambda}\}=0,\ee
 \be
\frac{[4F_{\mu\nu}-N_{\mu}N^{\lambda}(F_{\lambda\nu}+3\Omega_{\lambda\nu})+N_{\nu}N^{\lambda}(F_{\lambda\mu}-\Omega_{\mu\lambda}
)]\partial^{\mu}(\sqrt{g}\phi)}{\sqrt{g}\phi}$$$$
+\nabla^{\mu}[4F_{\mu\nu}-N_{\mu}N^{\lambda}(F_{\lambda\nu}+3\Omega_{\lambda\nu})+N_{\nu}N^{\lambda}(F_{\lambda\mu}-\Omega_{\mu\lambda}
)]$$$$+N_{\mu}(F_{\nu\lambda}+3\Omega_{\nu\lambda})\nabla^{\mu}N^{\lambda}+N^{\lambda}
(F_{\lambda\mu}+3\Omega_{\lambda\mu})\nabla_{\nu}N^{\mu}-N_{\lambda}(F_{\nu\mu}-\Omega_{\mu\nu})
\nabla^{\mu}N^{\lambda}$$$$-N^{\lambda}(F_{\lambda\mu}-\Omega_{\mu\lambda})\nabla^{\mu}N_{\nu}+2N^{\mu}R_{\mu\nu}-\frac{2\omega
N^{\mu}\partial_{\mu}\phi\partial_{\nu}\phi}
{\phi^2}-\frac{\zeta(x^{\alpha})N_{\nu}}{\phi}=0\ee and \be
G_{\mu\nu}=-\frac{8\pi}{\phi}T_{\mu\nu}+\frac{\omega\partial_{\mu}\phi\partial_{\nu}\phi}{\phi^2}+
\frac{\partial_{\mu}(\sqrt{g}\partial_{\nu}\phi)}{\sqrt{g}\phi}
-\frac{\zeta(x^{\alpha})N_{\mu}N_{\nu}}{\phi} +\frac{2\Box(\phi
N_{\mu}N_{\nu})}{\phi}\ee
$$-\frac{g_{\mu\nu}}{2\phi}\{2\Box\phi+\frac{\omega g^{\alpha\beta}\partial_{\alpha}\phi\partial_{\beta}\phi}{\phi}
-2\phi F_{\alpha\beta}F^{\alpha\beta}+2\phi
N_{\alpha}N^{\beta}F^{\alpha\lambda}\Omega_{\beta\lambda}$$$$+\phi
N_{\alpha}N^{\beta}(F^{\alpha\lambda}F_{\beta\lambda}+\Omega^{\alpha\lambda}\Omega_{\beta\lambda})+2N^{\alpha}N^{\beta}(\phi
R_{\alpha\beta}-\frac{\omega\partial_{\alpha}\phi\partial_{\beta}\phi}{\phi})\}+
\frac{V(\phi)}{\phi} g_{\mu\nu}.$$  where we defined \be
\Box=\frac{1}{\sqrt{g}}\partial_{\mu}(\sqrt{g}g^{\mu\nu}\partial_{\nu}).\ee
We now choose flat FRW background metric to study stability
situations of $\Lambda CDM$ vacuum de Sitter, dust and radiation
eras of the model.
\section{Cosmological setting}
  In context of homogenous and isotropic universes, one use usually FRW background
  metric which from point of view of a comoving observer, in flat case with Lorentizan signature $(-,+,+,+)$ is given by
   \begin{equation} ds^2=-dt^2+a(t)^2(dx^2+dy^2+dz^2).\end{equation} $a(t)$ is scale factor of spatial part of the above metric.
    Applying (3.1) one can obtain a simple solution of
   the equation (2.9) described by Kronecker delta function as \begin{equation} N_{\mu}(t)=\delta_{t\mu}\end{equation}
    where $N_{\mu}=1$ for $\mu=t$ and $N_{\mu}=0$ for $\mu\neq t.$ Applying (3.2) and definition of covariant differentiation
    $\nabla\equiv\partial+\Gamma$  for
   (2.4) one can obtain  \begin{equation}
F_{\mu\nu}(t)=0,~~~~~\Omega_{\mu\nu}(t)=4a\dot{a}~diag(0,1,1,1)\end{equation}
where  $\dot~\equiv\frac{d}{dt}.$ Applying (3.1), (3.2) and (3.3),
the equations (2.10) and (2.11) become respectively
 \begin{equation}
6\omega\dot{\psi}+12\dot{H}+30\omega H\psi +5\omega
\psi^2+18H^2+\frac{dV(\phi)}{d\phi}=0\end{equation} and \be
\frac{\zeta}{\phi}=2\omega\psi^2-6(\dot{H}+H^2)\ee where we
defined \be \psi(t)=\frac{\dot{\phi}}{\phi},\ee and \be
H(t)=\frac{\dot{a}}{a}.\ee Inserting (2.8), (3.1), (3.2), (3.3),
(3.5), (3.6) and (3.7) one can obtain time-time and space-space
components of the Einstein equation (2.12) respectively as
follows.
\begin{equation}G^t_t=3H^2=8\pi \rho^*,\end{equation}
and\begin{equation}G^i_j=(2\dot{H}+3H^2)\delta^i_j=-8\pi
p^*\delta^i_j\end{equation} where  $\delta^i_j$ with
$i,j\equiv\{x,y,z\}$ is 3 dimensional Keonecker delta function.
Also we defined generalized fluid density $\rho^*$ and
corresponding isotropic pressure $p^*$ as
\begin{equation}
8\pi\rho^*=\frac{8\pi(\rho_m+\rho_r)}{\phi}+\frac{(5\omega+4)}{2}\psi^2+
2\dot{\psi}-9\dot{H}+6H\psi-9H^2+\frac{V(\phi)}{\phi}
\end{equation}and \begin{equation}8\pi p^*=\frac{8\pi \rho_r}{3\phi}-\frac{(2-\omega)\psi^2}{2}-\dot{\psi}+3\dot{H}-3H\psi+3H^2-
\frac{V(\phi)}{\phi}\end{equation} where $(\rho_m\neq0, p_m=0)$
and $(\rho_r,p_r)\neq0$ with $p_r=\rho_r/3$ are matter and
radiation components of the mixture perfect fluid with total
density $\rho=\rho_m+\rho_r$ and pressure $p=p_r.$ Applying (3.8)
and (3.9),
 the Bianchi identity
$\nabla_{\mu}G^{\mu}_{\nu}=0$ leads to covariant conversation
condition
\begin{equation}\dot{\rho}^*+3H(\rho^*+p^*)=0.\end{equation}
Inserting (3.8), the above conservation condition can be rewritten
as \be \frac{p^*}{\rho^*}=-1-\frac{2}{3}\frac{\dot{H}}{H^2}\ee
which can be re-derived directly from (3.8) and (3.9). However one
can eliminate $\dot{H}$ term of the equation (3.5) by inserting
(3.13) to obtain barotropic parameter of the effective fluid as
\be
\gamma=\frac{p^*}{\rho^*}=-\frac{1}{3}-\frac{2\omega\psi^2}{9H^2}+\frac{\zeta}{9\phi
H^2}.\ee The above equation shows that the fields
$\zeta(t),\psi(t)$ and $H$ can be control numerical values of
$\gamma.$ For instance $\psi=\zeta=0$ reads to cosmic strings
$\gamma=-\frac{1}{3}.$ In what follows we will seek stability of
phase solutions for de Sitter, dust and radiation eras by setting
ansatz $\gamma=-1,0,\frac{1}{3}$ respectively. One can obtain a
good constraint condition between relative densities counterparts
as \be12(1+2\omega)\frac{
V(\phi)}{\phi}+\frac{dV(\phi)}{d\phi}-6(2+5\omega)\bigg(\frac{8\pi\rho_m}{\phi}\bigg)-4(5+12\omega)\bigg(\frac{8\pi\rho_r}{\phi}\bigg)$$$$
+12\omega H\psi-\omega(43+102\omega)\psi^2+18(1+2\omega)H^2=0\ee
where matter and radiation densities counterparts $\rho_{m,r}$
satisfy separately the conservation equation respectively as
follows. \be \dot{\rho}_m+3H\rho_m=0\ee and
\be\dot{\rho}_r+4H\rho_r=0.\ee The  condition (3.15) is obtained
from (3.11) when we eliminate $\dot{\psi}, p^*, \rho^*, \dot{H}$
via (3.4), (3.13), (3.8) and (3.5) respectively. Also we can
obtain a suitable equation for $\psi$ by applying (3.8), (3.9),
(3.10) and (3.11) such that
\be\dot{\psi}=-(1+17\omega)\psi^2+6H^2-3H\psi-8\bigg(\frac{8\pi\rho_r}{\phi}\bigg)-5\bigg(\frac{8\pi\rho_m}{\phi}\bigg)+4\frac{V(\phi)}{\phi}.
\ee We now can solve the equations (3.14), (3.15), (3.16), (3.17)
and (3.18) to determine the fields $a, H, \phi, \psi,  \zeta$ for
given sources  $\rho_m, \rho_r,$ and $V(\phi)$ via dynamical
system approach. To do so we must be  first make 5 dimensionless
phase space variables from $a, H, \phi, \psi, \zeta$ and then
obtain corresponding dynamical equations of phase space. To study
stability of de Sitter epoch we must be evaluate critical points
of phase space, and eigenvalues of corresponding Jacobi matrix as
follows.
\subsection{Cosmic dynamical system phase space} First we define
dimensionless time derivative against e-folding parameter
$\tau=\ln(a/a_i)$ of the expanding Universe as \be
^{\prime}=\frac{d}{d\tau}=\frac{1}{H}\frac{d}{dt}\ee together with
the following dimensionless variables of the cosmic phase space.
\be x(\tau)=\frac{\psi}{H},\ee
\begin{equation}q(\tau)=\frac{\zeta}{\phi H^2},\ee
\be y(\tau)=\frac{8\pi\rho_m}{\phi H^2}\ee \be
z(\tau)=\frac{8\pi\rho_r}{\phi H^2}\ee \be v(\tau)=\frac{
V(\phi)}{\phi H^2}\ee and \be s(\tau)=\frac{\dot{H}}{H^2}.\ee
Inserting (3.19), (3.20), (3.21), (3.22), (3.23), (3.24), (3.25)
into the equations (3.14), (3.15), (3.16), (3.17) and (3.18), one
can obtain dimensionless dynamical equations of phase space
variables as follows. \be
x^{\prime}=-(1+17\omega)x^2+\frac{3(\gamma-1)}{2}x-5y-8z+4v+108\ee
 \be
y^{\prime}=(3\gamma-2-x)y,\ee \be z^{\prime}=[3(\gamma-1)-x]z,\ee
\be v^{\prime}=\omega(43+102\omega)x^3-12\omega
x^2+18(1+2\omega)x+6(2+5\omega)xy$$$$
+4(5+12\omega)xz-(13+24\omega)xv+\frac{3(1+\gamma)v}{2}\ee where
\be q=2\omega x^2+3(1+3\gamma),\ee \be s=-\frac{3}{2}(1+\gamma)\ee
and we used \be \frac{
dV(\phi)}{d\phi}=H^2\bigg(\frac{v^{\prime}}{x}+v+\frac{2sv}{x}\bigg).\ee
The equations (3.26) to (3.29) describe dynamical equations of a
4D phase space $\{x,y,z,v\}.$ They are first order nonlinear
differential equations and so their solutions may have choatic
behavior near possible critical points. If we want to seek
stabiliy of phase solutions of the above dynamical equations, then
we must be calculate their possible critical points for vacuum de
Sitter era by setting $\gamma=-1.$ Next we obtain eigenvalues of
the corresponding Jacobi matrix and discuss their characteristics
(see table 1).
\subsection{ $\Lambda CDM$ de Sitter era}
Inserting $\gamma=-1$  the dynamical equations (3.26), (3.27),
(3.28), (3.29) can be rewritten as \be
x^{\prime}=-(1+17\omega)x^2-3x-5y-8z+4v+108\ee
 \be
y^{\prime}=-(5+x)y,\ee \be z^{\prime}=-(6+x)z,\ee \be
v^{\prime}=\omega(43+102\omega)x^3-12\omega
x^2+18(1+2\omega)x+6(2+5\omega)xy$$$$
+4(5+12\omega)xz-(13+24\omega)xv\ee where (3.30) and (3.31) take
the following forms respectively. \be q=2\omega x^2-6,\ee and \be
s=0.\ee
  For the vacuum de Sitter era,
matter and radiation densities counterparts are negilible and so
we must be set \be y=0,~~~z=0.\ee $x,v$ components of the critical
points are determined by solving $x^{\prime}=0=v^{\prime}.$
Inserting (3.39) and using (3.33) and (3.36) the equations
$x^{\prime}=0$ and $v^{\prime}=0$ become
\be(1+17\omega)x_c^2+3x_c-(4v_c+108) = 0\ee and
\be\omega(43+102\omega)x_c^3-12\omega
x_c^2+18(1+2\omega)x_c-(13+24\omega)x_cv_c = 0.\ee
$(x_c=0,v_c=-27)$ satisfyes trivially the equations (3.40) and
(3.41) for arbitrary values of $\omega$ and so it is one of de
Sitter era critical points.  If $x_{c}\neq0$ then the equation
(3.41) become \be\omega(43+102\omega)x_c^2-12\omega
x_c+18(1+2\omega)-(13+24\omega)v_c = 0.\ee
 Eliminating $v_c$ between (3.40)
and (3.42), we obtain \be
(73\omega+13)x^2+(120\omega+39)x-(2736\omega+1476)=0\ee which has
two solutions as \be
x_c^{\pm}=\frac{-3(40\omega+13)\pm\sqrt{90368\omega^2+64736\omega+8697}}{2(73\omega+13)}.\ee
Eliminating $\omega$ between (3.40) and (3.43) we obtain \be
 v^{\pm}_c(x_c^{\pm})=\frac{3(49x_c^4+167x_c^3-5892x_c^2-3528x_c+49248)}{2(73x_c^2+120x_c-2736)}\ee
in which $x_c$ must be inserted from (3.44). The solutions (3.44)
and (3.45) show that there is
  two class of fixed points as \be P_{2}^{de~Sitter}(\omega):(x_c^{+}(\omega),y_c=0,z_c=0,v_c^+(\omega))\ee and \be P_3^{de~Sitter}(\omega):(x_c^-(\omega),y_c=0,z_c=0,v_c^+(\omega))\ee
  which make infinite number of critical points against different values of $\omega.$  Setting $x^+_c=0$ we
obtain $\omega=-0.16856$  where $P_2^{de~Sitter}$ reaches to
$P_1^{de~Sitter}$ and they become a unique fixed point. If we
choose $x_c^-=0$ we obtain $\omega=-0.56038$ where
$P_3^{de~Sitter}$ and $P_1^{de~Sitter}$ become a unique fixed
point. They have stable behavior for $\omega<0$ and saddle
(quasi-stable) for $\omega\geq0$ (see figure 1 and table 1).
Setting $x_c^+=x_c^-$ we obtain $\omega=-0.17915$ where
$P_{2,3}^{de~Sitter}$ describes a quansi-stable state (see table
1). In general relativity approach of the BD theory itself we know
$\omega\to+\infty$ where the BD scalar field reaches to a constant
value. Hence we choose also samples $\omega=40000$ and
$\omega=-40000$ to obtain numerical values of critical points
components as follows.
 \be P^{de Sitter}_1( \forall\omega\in\mathbb{R}):
(x_c=0,y_c=0,z_c=0,v_c=-27),\ee \be
 P_{1,2}^{de~Sitter}(\omega=-0.16856):(x_c=0,y_c=0,z_c=0,v_c=-27),\ee
 \be
 P_3^{de~Sitter}(\omega=-0.16856):(x_c=-27.02,y_c=0,z_c=0,v_c=591.90),\ee
\be
 P_{1,3}^{de~Sitter}(\omega=-0.56038):(x_c=0,y_c=0,z_c=0,v_c=-27),\ee
 \be
 P_2^{de~Sitter}(\omega=-0.56038):(x_c=-1.013,y_c=0,z_c=0,v_c=-25.15),\ee
\be
 P_{2,3}^{de~Sitter}(\omega=-0.17915):(x_c=112.92,y_c=0,z_c=0,v_c=12953.62),\ee
\be
 P_{2}^{de~Sitter}(\omega=40000):(x_c=1.24,y_c=0,z_c=0,v_c=-21.99),\ee
\be
 P_{3}^{de~Sitter}(\omega=40000):(x_c=-2.88,y_c=0,z_c=0,v_c=-5.99),\ee
\be
 P_{2}^{de~Sitter}(\omega=-40000):(x_c=-2.88,y_c=0,z_c=0,v_c=-5.99),\ee
\be
 P_{3}^{de~Sitter}(\omega=-40000):(x_c=1.24,y_c=0,z_c=0,v_c=-21.99).\ee
 Other  critical fixed points which
can be considerable physically is for situations where at least
one of roots of second order equations (3.40) and (3.42) have
similar value (common root). To do so we must be set the following
constriant condition between their coefficients.\be
 \frac{(1+17\omega)}{\omega(43+102\omega)}=-\frac{3}{12\omega}=\frac{4v_c+108}{(13+24\omega)v_c-18(1+2\omega)}\ee
leading to the following particular values. \be
\omega=\frac{47}{170}=0.27647,~~~v_c=68.870.\ee
 Inserting (3.59) the equations
(3.40) and (3.43) read $x_c^{+}=7.9433,~~x_c^-=-8.4697$ and so we
will have two other critcal fixed points more as follows.
 \be
 P^{de
Sitter}_{2}(\omega=0.27647):(x_c=7.9433,y_c=0,z_c=0,v_c=68.87)\ee
 and\be
 P^{de
Sitter}_{3}(\omega=0.27647):(x_c=-8.4697,y_c=0,z_c=0,v_c=68.87).\ee
   where nature of the fixed point (3.60) is stable but for (3.61) is unstable respectively (see table 1 and figure 1).
  Stability and/or instability
of the above critical points can be follow via arrow diagrams of
the dynamical equations (3.33) to (3.36) in figure 1 against
different values of $\omega.$  In general, we can obtain time
dependent solutions of the field equations of $\Lambda CDM$ era
for critical points $P_{1,2,3}^{de~Sitter}(\omega)$ as follows.
  \be  P_{1,2,3}^{de~Sitter}:\left(%
\begin{array}{c}
  \frac{\phi(t)}{\phi_0}=e^{x_cHt} \\
   \rho_m=0 \\
  \rho_r=0\\
   V(\phi)=v_cH^2\phi \\
  \zeta(t)=(2\omega x_c^2-6)\phi_0H^2e^{x_cHt} \\
   \frac{a(t)}{a_0}=e^{Ht} \\
\end{array}%
\right)\ee  where $H$ is Hubble constant which must be inserted
via observational data and numerical values of $(x_c,v_c)$ should
be inserted from the equations (3.46) to (3.57) and/or (3.60) to
(3.61).  If we want to determine which of the above critical
points have stable behavior then we must be calculate
corresponding Jacobi matrix (1.4) and obtain eignevalues as
follows (see table 1).
\be  J^{de~Sitter}_{1,2,3}(\omega)=$$$$\left(%
\begin{array}{cccc}
   -3-2(1+17\omega)x_c & -5 & -8 & 4 \\
  0 &  -(5+x_c) & 0 & 0 \\
  0 & 0 &  -(6+x_c) & 0 \\
  F(\omega,x_c,v_c) & 6(2+5\omega)x_c &  4(5+12\omega)x_c &  -(13+24\omega)x_c \\
\end{array}%
\right)\ee  where we defined\be
 F(\omega,x_c,v_c)=3\omega(43+102\omega)x_c^2-14\omega
x_c+18(1+2\omega)-(13+24\omega)v_c\ee
  and numerical values of
$\omega,x_c,v_c$ should be inserted from the equations (3.48) to
(3.57) and/or (3.60) to (3.61). We obtain corresponding secular
equation as \be
 (\lambda+5+x_c)(\lambda+6+x_c)[\lambda^2+[3+(15+58\omega)x_c]\lambda
$$$$ +(13+24\omega)[3x_c+2(1+17\omega)x_c^2]]=0\ee
which has four eigenvalues as \be
\lambda_1=-(5+x_c),~~~\lambda_2=-(6+x_c),$$$$ \lambda_{3}=
-\frac{[3+(15+58\omega)x_c]}{2}$$$$
 +\frac{1}{2}\sqrt{[3+(15+58\omega)x_c]^2-4(13+24\omega)x_c[3+2(1+17\omega)x_c]}$$$$
\lambda_{4}= -\frac{[3+(15+58\omega)x_c]}{2}$$$$
 -\frac{1}{2}\sqrt{[3+(15+58\omega)x_c]^2-4(13+24\omega)x_c[3+2(1+17\omega)x_c]}\ee
 where  $\lambda_{1,2,3,4}<0$ and
$\lambda_{1,2,3,4}>0$ describ stable and unstable state of the
system. If some of the eigenvalues take positive values
numerically but some other ones become negative then the system
will be take quasi stable state namely saddle (see figure 1). We
insert numerical values of $(\omega,x_c)$ from the equations
(3.48) to (3.57) and/or (3.60) to (3.61) and collect numerical
values of
 eigenvalues $\lambda_{1,2,3,4}^{de~Sitter}$ in table 1 where  first column in right side denotes to their stability and/or instabiity  nature.
   As a result of our work we now study experimental
correspondence of our obtined solutions.
 Correspondence between Newton`s
gravity coupling parameter and the BD scalar field is well known
as $\phi\equiv\frac{1}{G}$ from the BD gravity theory which by
inserting (3.6)  one infers [1] \be
\frac{1}{H}\bigg|\frac{\dot{G}}{G}\bigg|_{SBD}=\frac{1}{|1+\omega|}\ee
 while for our model we will have
\be
 \frac{1}{H}\bigg|\frac{\dot{G}}{G}\bigg|^{\pm}_{VBD}=\bigg|
\frac{-3(40\omega+13)\pm\sqrt{90368\omega^2+64736\omega+8697}}{2(73\omega+13)}\bigg|\ee
  which in GR limits
$\omega\to+\infty$ we can obtain nonzero counterpart of preferred
reference frame effects as follows.\be
 \lim_{\omega\to+\infty}\frac{1}{H}\bigg|\frac{\dot{G}}{G}\bigg|^{\pm}_{VBD}-
\lim_{\omega\to+\infty}\frac{1}{H}\bigg|\frac{\dot{G}}{G}\bigg|_{SBD}\approx\bigg|^{1.24;for
+}_{2.88;for -} \ee  where the present value of the Hubble
constant is [1](see also [10,29]) \be
H_{obs}=7.24\times10^{-11}~yr^{-1}.\ee
  The above result predicts
non-valishing $\dot{G}$ in presense of dynamical vector fields
effects even in GR limits $\omega>>1$ which in BD gravity itself
can not be detected. We now study dust era and its stability
conditions of
 our model in the following subsection.
\subsection{ Dust era}  For dust era matter density is non-vanishing $y\neq0$ but for
the radiation density we have $z=0$ and corresponding barotropic
index is $\gamma=0.$ Using the latter initial conditions the
dynamical equations (3.26), (3.27), (3.28), (3.29) read \be
x^{\prime}=-(1+17\omega)x^2- 3x/2-5y+4v+108,\ee \be
y^{\prime}=-(2+x)y,\ee \be z^{\prime}=0\ee \be
v^{\prime}=\omega(43+102\omega)x^3-12\omega
x^2+18(1+2\omega)x+6(2+5\omega)xy-(13+24\omega)xv+ 3v/2,\ee
 and (3.30) and (3.31) become
respectively  \be q= 3+2\omega x^2\ee and \be s=-\frac{3}{2}.\ee
Critical points are obtained by
 using (3.71) to (3.74) and setting
$x^{\prime}=0=y^{\prime}=v^{\prime}$ as \be x_{ c}=-2,~~~
y_c(\omega)=\frac{10244\omega+6173}{83},~~~v_c(\omega)=
\frac{14216\omega+5496681}{83}\ee where
 $\omega>-0.6026$ because of
positivity condition $y>0$ of the matter density (3.22).
 However one can infers that $\omega$ dependent
single critical point in the dust era become \be
P^{Dust}:\bigg(x_{ c}=-2, y_c=\frac{10244\omega+6173}{83},
z_{c}=0, v_c=\frac{14216\omega+5496681}{83}\bigg)\ee
 where (3.75) become \be
q_c=3+8\omega.\ee   Setting
$\omega=\{-0.16856,-0.56038,-017915,40000,0.27647\}$ the above
dust era critical point become respectively \be
 P^{Dust}(\omega=-0.16856):\bigg(x_c=-2,y_c=53.57,z_c=0,v_c=67830.68\bigg)\ee
\be
 P^{Dust}(\omega=-0.56038):\bigg(x_c=-2,y_c=5.21,z_c=0,v_c=67761.90\bigg),\ee
\be
 P^{Dust}(\omega=-0.17915):\bigg(x_c=-2,y_c=52.26,z_c=0,v_c=67828.82\bigg)\ee
 \be
P^{Dust}(\omega=40000):\bigg(x_c=-2,y_c=4.94\times10^6,z_c=0,v_c=7.09\times10^6\bigg)\ee
\be
 P^{Dust}(\omega=0.27647):\bigg(x_c=-2,y_c=108.50,z_c=0,v_c=67908.78\bigg).\ee
One can calculate Jacobi matrix (1.4) for the critical point
(3.78) as follows. \be
J^{Dust}(\omega)=\left(%
\begin{array}{cccc}
  \frac{(5+136\omega)}{2} &  -5 &  0 &  4\\
   -\frac{(10244\omega+6173)}{83} &  0 &  0 & 0\\
 0 & 0 & 0 & 0 \\
   -\frac{(239592\omega^2+132055352\omega+71455359)}{83} &  -12(2+5\omega) &  0 & \frac{(55+96\omega)}{2} \\
\end{array}%
\right)\ee  where its secular equation defined by (1.5) become \be
 \lambda\big[\lambda^3-(30+116\omega)\lambda^2$$$$ +
(\frac{1143185109}{332}+\frac{528335358\omega}{83}+\frac{1229280\omega^2}{83})
\lambda+\frac{6173}{2}+5122\omega\big]=0.\ee
 Inserting
$\omega=\{-0.16856,-0.56038,-017915,40000,0.27647\}$ we obtain
numerical solutions of the eigenvalues equation (3.86) for
critical points (3.80) to (3.84) and collect them into the table
1.
   Inserting (3.78) into the equations (3.20) to (3.25) and some simple
  integral calculations
one can obtain dust era solutions as follows.  \be
P^{Dust}(\omega):\left(%
\begin{array}{c}
   \frac{\phi(t)}{\phi_0}=\big(\frac{t}{t_0}\big)^{\frac{4}{3}} \\
   \rho_m(t)=\frac{(10244\omega+6173)}{747\pi }\frac{\phi0}{t_0^2}\big(\frac{t}{t_0}\big)^{-\frac{2}{3}} \\
   \rho_r(t)=0 \\
  V(\phi)=\big(\frac{16864\omega}{747}+\frac{7328908}{249}\big)\frac{\phi_0}{t_0^2}\big(\frac{\phi}{\phi_0}\big)^{-\frac{1}{2}} \\
   \zeta(t)=\frac{4(3+8\omega)}{9}\frac{\phi_0}{t_0^2}\big(\frac{t}{t_0}\big)^{-\frac{2}{3}} \\
 \frac{a(t)}{a_0}=\big(\frac{t}{t_0}\big)^{\frac{2}{3}} \\
\end{array}%
\right)\ee  where $\phi_0=\phi(t_0),a_0=a(t_0),$ and $t_0$ is an
arbitrary constant time. In the following subsection we study
radiation era of  the model and its stability conditions.
\subsection{Radiation era}  In case of radiation era, the matter density
is vanishing $y=0$ and barotropic index  of state equation of
radiation is $\gamma=\frac{1}{3}.$ Inserting the latter initial
conditions the dynamical equations (3.26), (3.27), (3.28), (3.29)
read \be x^{\prime}=-(1+17\omega)x^2-x-8z+4v+108,\ee \be
y^{\prime}=0,\ee \be z^{\prime}=-(2+x)z\ee \be
v^{\prime}=\omega(43+102\omega)x^3-12\omega
x^2+18(1+2\omega)x+4(5+12\omega)zx-(13+24\omega)xv+2v,\ee
 where (3.30) and (3.31) become
respectively \be q=6+2\omega x^2\ee and \be s=-2.\ee Critical
points are obtained  by using (3.88) to (3.91) and setting
$x^{\prime}=0=z^{\prime}=v^{\prime}$ as \be
 x_c=-2,~~~ v_{c}=\frac{283}{4}+\frac{349\omega}{2},~~~
 z_c=\frac{315\omega}{4}+\frac{389}{8}\ee
where positivity condition of the radiation  density (3.23)
restricts us to choose $z>0$  and so $\omega>-0.59846.$ Thus
critical point in the radiation era become \be P^{Radiation}:(x_{
c}=-2,y_{c}=0,
 z_c=\frac{315\omega}{4}+\frac{389}{8}, v_{c}=\frac{283}{4}+\frac{349\omega}{2}
)\ee  where \be q_{c}=6+8\omega,~~~ s_{c}=-2.\ee
 Setting
$\omega=\{-0.16856,-0.56038,-017915,40000,0.27647\}$ the above
radiation era critical point become respectively \be
 P^{Rad}(\omega=-0.16856):\bigg(x_c=-2,y_c=0,z_c=35.35,v_c=41.34\bigg)\ee
\be
 P^{Rad}(\omega=-0.56038):\bigg(x_c=-2,y_c=0,z_c=4.50,v_c=-27.04\bigg),\ee
\be
 P^{Rad}(\omega=-0.17915):\bigg(x_c=-2,y_c=0,z_c=34.52,v_c=39.49\bigg)\ee
 \be
 P^{Rad}(\omega=40000):\bigg(x_c=-2,y_c=0,z_c=3.15\times10^6,v_c=6.98\times10^6\bigg)\ee
\be
 P^{Rad}(\omega=0.27647):\bigg(x_c=-2,y_c=0,z_c=70.40,v_c=118.99\bigg).\ee
One can calculate Jacobi matrix (1.4) for the radiation era
critical point (3.95) as follows. \be
J^{radiation}(\omega)=\left(%
\begin{array}{cccc}
  3+68\omega &  0 &  -8 & 4\\
  0 &  0 &  0 & 0\\
  -\big(\frac{315\omega}{4}+\frac{389}{8}\big) & 0 & 0 &0\\
   -1632\omega^2-\frac{979\omega}{2}+\frac{283}{4} &  0&  -8(5+12\omega) & 4(7+12\omega)\\
\end{array}%
\right)\ee where its secular equation (1.5) become \be
 \lambda[\lambda^3-(116\omega+31)\lambda^2+(9792\omega^2+3376\omega-588)\lambda+5040\omega+3112]=0.\ee
 Inserting
$\omega=\{-0.16856,-0.56038,-017915,40000,0.27647\}$ we solve
(3.103) and obtain numerical values of eigenvalues for critical
points (3.97) to (3.101) and collect them into the table 1.
   Inserting (3.95) and (3.96) into the equations (3.20) to (3.25) and some simple
  integral calculations
one finds   \be
 P^{Rad}(\omega):\left(%
\begin{array}{c}
   \frac{\phi(t)}{\phi_0}=\frac{t}{t_0} \\
   \rho_m(t)=0 \\
  \rho_r(t)=\frac{1}{128\pi}\big(315\omega+\frac{389}{2}\big)\frac{\phi_0}{t_0^2}\big(\frac{t}{t_0}\big)^{-1} \\
   V(\phi)=\big(\frac{349\omega}{8}+\frac{283}{16}\big)\frac{\phi_0}{t_0^2}\big(\frac{\phi}{\phi_0}\big)^{-1}\\
   \zeta(t)=\frac{(3+4\omega)}{2}\frac{\phi_0}{t_0^2}\big(\frac{t}{t_0}\big)^{-1}\\
  \frac{a(t)}{a_0}=\big(\frac{t}{t_0}\big)^{\frac{1}{2}} \\
\end{array}%
\right)\ee  where $\phi_0=\phi(t_0),a_0=a(t_0),$ and $t_0$ is an
arbitrary constant time.
\begin{center}
\begin{tabular}{|c|c|c|}
  \hline
 $Fixed~point$  & $Eignevalues: (\lambda_1,~\lambda_2,~\lambda_3,~\lambda_4)$ & $Nature$ \\
  \hline
  $P_{1;\forall\omega\in\mathbb{R}}^{deS}$ & (-5,~-6,~0,~-6) & $stable$ \\
\hline
  $P_{1,2;\omega_1=-0.16856}^{deS}$ & (-5,~-6,~0,~-6 )& $stable$ \\
  \hline
  $P_{3;\omega=-0.16856}^{deS}$ & (22.02,~21.02,~483.90,~-207.63)& $saddle$ \\
  \hline
  $P_{1,3;\omega=-0.56038}^{deS}$ & (-5,~-6,~0,~-6) & $stable$ \\
  \hline
  $P_{2;\omega=-0.56038}^{deS}$ & (-3.99,~-4.99,~-0.91,~-40.55) & $stable$ \\
  \hline
  $P_{2,3;\omega=-0.17915}^{deS}$ & (-117.92,~-118.90,~917.93,~-1964.90) & $saddle$ \\
  \hline
  $P_{2;\omega=40000}^{deS}$ & (-6.24,~-7.24,~$-2.38\times10^6$,~$-3.37\times10^6$) & $stable$\\
  \hline
   $P_{3;\omega=40000}^{deS}$ & (-2.12,~-3.12,~$7.83\times10^6$,~$5.53\times10^6$) & $saddle$ \\
  \hline
  $P_{2;\omega=-40000}^{deS}$ & (-2.12,~-3.12,~$-5.53\times10^6$,~$-7.83\times10^6$) & $stable$ \\
  \hline
 $P_{3;\omega=-40000}^{deS}$ & (-6.24,~-7.24,~$3.37\times10^6$,~$2.38\times10^6$)& $saddle$ \\
  \hline
  $P_{2;\omega=0.27647}^{deS}$ & (-12.94,~-13.94,~-187.11,~-311.44) & $stable$ \\
  \hline
   $P_{3;\omega=0.27647}^{deS}$ & (3.47,~2.37,~332.61,~187.11) & $untable$ \\
  \hline
  $P_{\omega=0.27647}^{Dus}$ & (0,~31.04+2281.09i,~-0.0009,~31.04-2281.09i) & $saddle$ \\
  \hline
  $P_{\omega=40000}^{Dus}$ & (0,~100,~$(2.3-4.3i)\times10^6$,~$(2.3+4.3i)\times10^6$)& $unstable$\\
  \hline
   $P_{\omega=-0.17915}^{Dus}$ & (0,~4.61+1517.70i,~-0.0009,~4.61-1517.70i) & $saddle$ \\
  \hline
 $P_{\omega=-0.56038}^{Dus}$ & (0,~0.002,~328.07,~-363.07) &$saddle$ \\
  \hline
$P_{\omega=-0.16856}^{Dus}$ & (0,~5.22+1539.73i,~-0.0009,~5.22-1539.73i) & $saddle$ \\
 \hline
  $P_{\omega=0.27647}^{Rad}$ & (0,~32.24+14.69i,~-3.41,~32.24-14.69i) & $saddle$ \\
  \hline
  $P_{\omega=40000}^{Rad}$ & (0,~0,~$(2.32-3.21i)\times10^6$,~$(2.32+3.21i)\times10^6$)& $unstable$ \\
  \hline
   $P_{\omega=-0.17915}^{Rad}$ & (0,~2.46,~34.09,~-26.33) & $saddle$ \\
  \hline
 $P_{\omega=-0.56038}^{Rad}$ & (0,~-16.75+17.26i,~-0.50,~-16.75-17.26i) & $stable$ \\
  \hline
$P_{\omega=-0.16856}^{Rad}$ & (0,~2.51,~34.82,~-25.88) & $saddle$ \\
  \hline
\end{tabular}
\end{center}
\begin{center}
 Table 1: Numerical values of
eigenvalues for $\Lambda CDM$ de Sitter, dust and radiation eras
where the corresponding space time scale factor become
$a_{deS}(t)\sim e^{Ht},$ $a_{Dus}(t)\sim t^{\frac{2}{3}}$ and
$a_{Rad}(t)\sim t^{\frac{1}{2}}$ respectively.
\end{center}
\section{Concluding remark} Applying VBD gravity [3] in presence of additional perfect fluid matter and self interaction
potential action     functionals we studied flat FRW space time
dynamics. We applied dynamical system approach
  to seek stable critical points
for vacuum de Sitter, dust and radiation eras. To do so we
calculate eigenvalues of the corresponding Jacobi matrix defined
on 4D phase space. In general, we obtain 3 type critical fixed
points for de Sitter era but 1 type for dust and radiation eras.
Nature of these critical points are depended to choose numerical
values of the BD parameter $\omega$. When the potential behaves as
(effective cosmological) constant then one of the critical fixed
point in de Sitter era become stable for $\omega<0$ and saddle for
$\omega\geq0.$ While for linear potential $V(\phi)\sim\phi$
(variable cosmological parameter) there is still a stable critical
point in de Sitter era but for particular value of
$\omega=0.27647.$ There is not obtained conditions where the all 3
fixed points reach to a unique critical fixed point. While  for
$\omega=\{-0.16856,-0.56038,-0.17915\}$ there is at least 2 out of
3 critical fixed points in de Sitter era which become unique (see
table 1).  In dust era the system become stable for
$\omega=-0.56038$ but behaves as unstable by vanishing matter
density for $\omega=0.27647$ (see figure 1). The latter case
predicts a phase transition from matter to vacuum de Sitter era.
Radiation era become quasi-stable for
$\omega=\{-0.56038,0.27647\}$ by vanishing the radiation density.
This result predicts a phase transition between radiation and dust
eras for particular value of $\omega=0.27647.$ Comparing diagrams
given in figure 1 we can understand $\omega_{uniqie}=0.27647$ is
important value for the BD parameter in the used gravity model [3]
where flat FRW space time tolerates a radiation era by supporting
potential $V(\phi)\sim\phi^{-1},$ then transmit to a dust era by
supporting a potential as $V(\phi)\sim\phi^{-\frac{1}{2}}$ and
finally transmit to a vacuum de Sitter era by supporting a linear
potential $V(\phi)\sim\phi.$ As a result of our work we consider
time dependent fluctuations of Newton`s coupling parameter
$\dot{G(t)}$ obtained from BD gravity itself and compare it with
our results in GR limits $\omega\to40000.$ Non-vanishing
counterparts denotes to preferred reference frame effects coming
from the used alternative model in this work. As extensions of our
work we seek preferred reference frame effects [3] on anisotropy
of Bianchi`s cosmology and galaxy rotation curves too in our next
work.
 \vskip 0.5cm
 {\bf\textbf{ References}}
\begin{description}
\item[1.] R. G. Salcedo, T. Gonzales and I. Quiros, Phys. Rev.
D92, 124056 (2015), gr-qc/1504.08315.
\item[2.] C. Brans, R. Dicke, Phys. Rev. 124, 925 (1961).
\item[3.] H. Ghaffarnejad, Gen. Rel.
Grav. 40, 2229 (2008); 41, 2941 (E) (2009).
\item [4.] V. G. Ivancevic and T. T. Ivancevic, \textit{Complex
Nonlinearity}, Chaos, Phase Transition, Topology Change and Path
Integrals, ( Springer verlag Berlin Heidelberg 2008).
\item[5.] J. M. T. Thompson and H. B. Stewart, \textit{Nonliniear Dynamics and
Chaos}, (Second Edition, John Wiley $\&$ Sons, LTD 2002).
\item[6.] K. Falconer \textit{Fractal Geometry}, Mathematical
foundations and applications, (Second Edition, John Wiley $\&$
Sons, 2003).
\item[7.] S. Y. Zhou, E. J. Copeland and P. M. Saffin, JCAP 0907, 009,
(2009); gr-qc/0903.4610 (2010).
\item[8.] T. Azizi and E. Yaraie, Int. J. Mod. Phys. D23, 2,
145002 (2014).
\item[9.] O. Hrycyna and M. Szydtowski, JCAP12,016 (2013); gr-qc/1310.1961.
\item[10.] O. Hrycyna, M. Szydlowski,  Phys. Rev. D 88 (2013) 6, 064018; gr-qc/1304.3300.
\item[11.] O. Hrycyna, M. Kamionka, M. Szydlowski, Phys. Rev. D 90 (2014) 12, 124040; astro-ph.CO/1404.7112.
\item[12.] E. J. Copeland, S. Mizuno, M. Shaeri, Phys. Rev. D 79 (2009) 103515, astro-ph.CO/0904.0877.
\item[13.] T. Matos, J. R. Luevano, I. Quiros, L. A. Urena-Lopez, J. A. Vazquez, Phys. Rev. D 80 (2009) 123521; astro-ph.CO/0906.0396.
\item[14.] L. A. Urena-Lopez, M. J. Reyes-Ibarra, Int. J. Mod. Phys. D 18 (2009) 621-634, astro-ph/0709.3996.
\item[15.] L. Amendola, Phys. Rev. D 62 (2000) 043511; astro-ph/9908023.
\item[16.] S. Fay, S. Nesseris and L. Perivolaropoulos, Phys. Rev. D 76, 063504
(2007);  gr-qc/0703006
\item[17.] K. Nozari and F. Kiani, Int. J. Geo. Meth. Mod. Phys.
8, 6, 1179 (2011).
\item [18.] G. Bertone, D. Hooper and J. Silk,
 Phys. Rept. 405, 279 (2005) hep-th/0404175.
\item [19.] J. A. Frieman,
     M. S. Turner and D. Huterer, Annu, Rev. Astron. Astrophys., 46, 385 (2008).
\item [20.] J. M. Moffat,
J. Cosmol. Astropart. Phys.03, 004 (2006), gr-qc/0506021v7.
\item [21.] J. D. Bekensten, Phil. Trans. R. Soc. A369, 5003 (2011); astro-ph.CO/1201.2759v1 (2012).
\item [22.] X. M. Deng, Y.Xie and T. Y. Huang,  Phys. Rev. D79, 044014, gr-qc/0901.3730v1 (2009).
\item[23.] H. Ghaffarnejad, Class. Quant. Grav. 27, 015008
(2010).
\item[24.] H. Ghaffarnejad, Journal of Physics: Conference series,
633, 012020 (2015).
\item[25.] \textit{Theory And Experiment In Gravitational
Physics}, (C. M. Will, Cambridge University press (1993)); revised
version: gr-qc/9811036.
\item[26.] E. Gaztanaga and J. A. Lobo, Astrophys. J., 548, 47
(2001).
\item[27.] R. D. Reasenberg et al, Astrophys. J., 234, 925 (1961).
\item[28.] C. M. Will, Living Rev. Rel. 9 (2006); http://WWW.livingreviews.org/lrr-2006-3.
\item [29.] B. Bertotti, L. Iess and P. Trotora, Nature (London),
425, 374 (2003); J. P. Uzan, astro-ph/0409424
\end{description}
\begin{figure}[ht!]
\centering
\includegraphics[width=1.6in,height=1.6in]{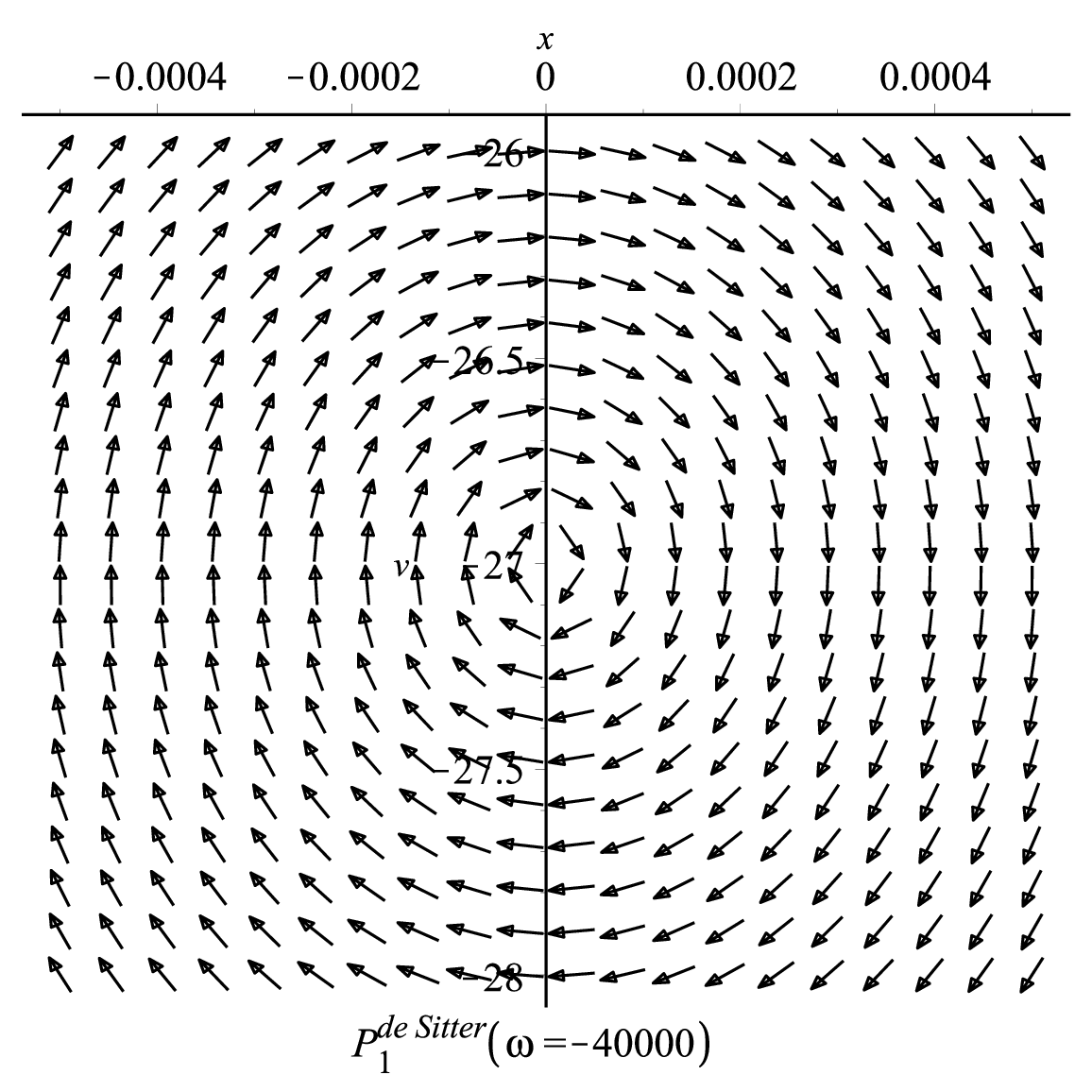}
\includegraphics[width=1.6in,height=1.6in]{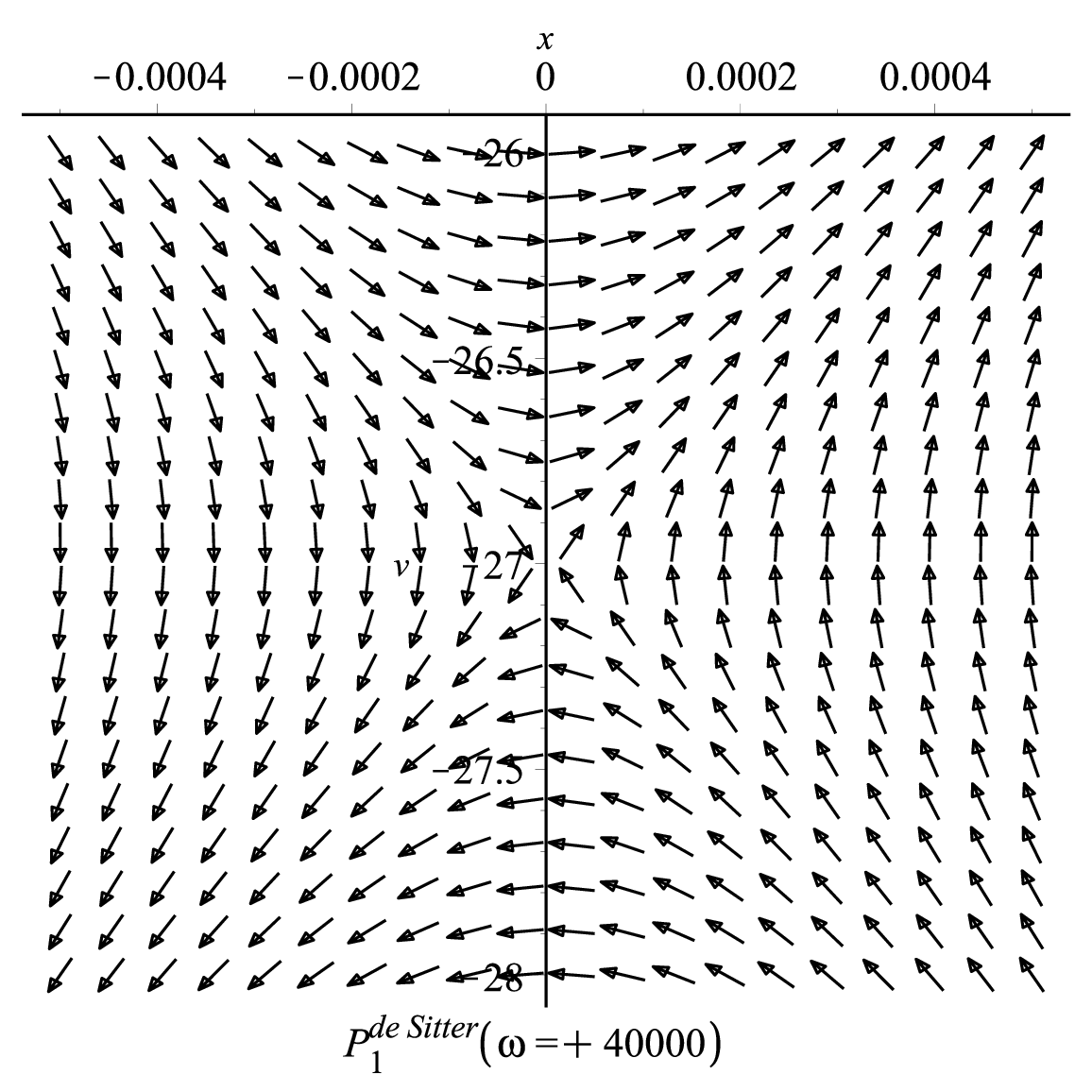}
\includegraphics[width=1.6in,height=1.6in]{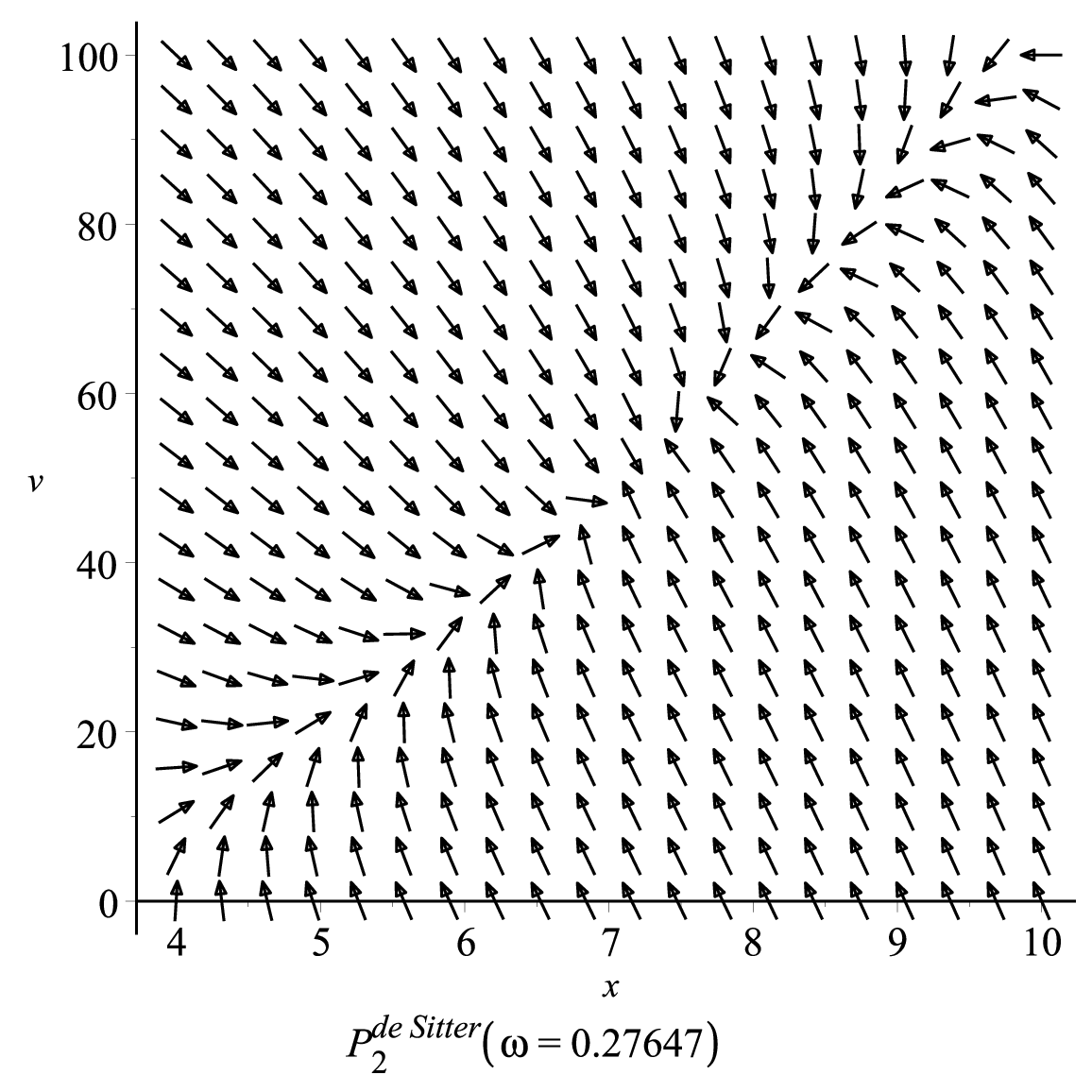}
\includegraphics[width=1.6in,height=1.6in]{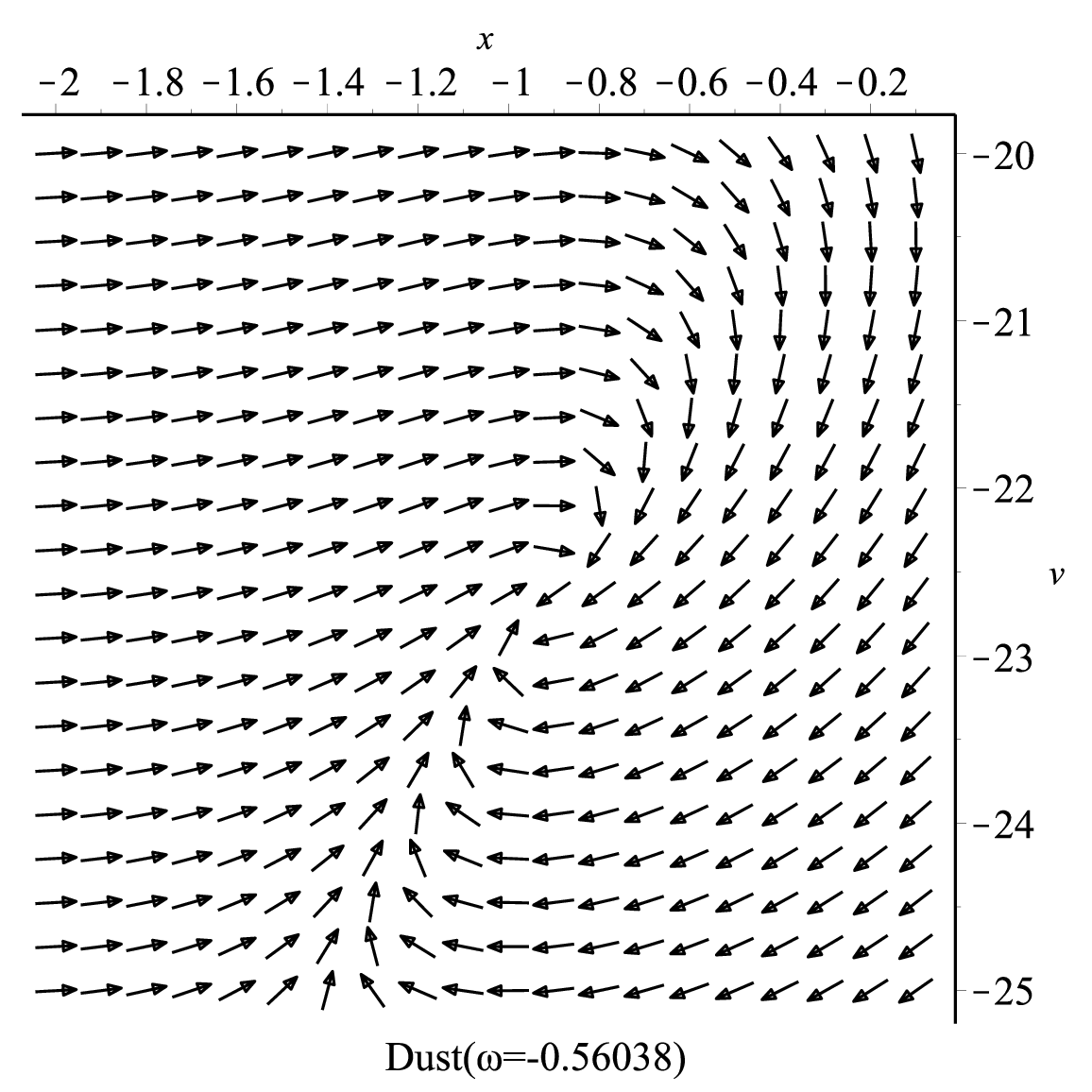}
\includegraphics[width=1.6in,height=1.6in]{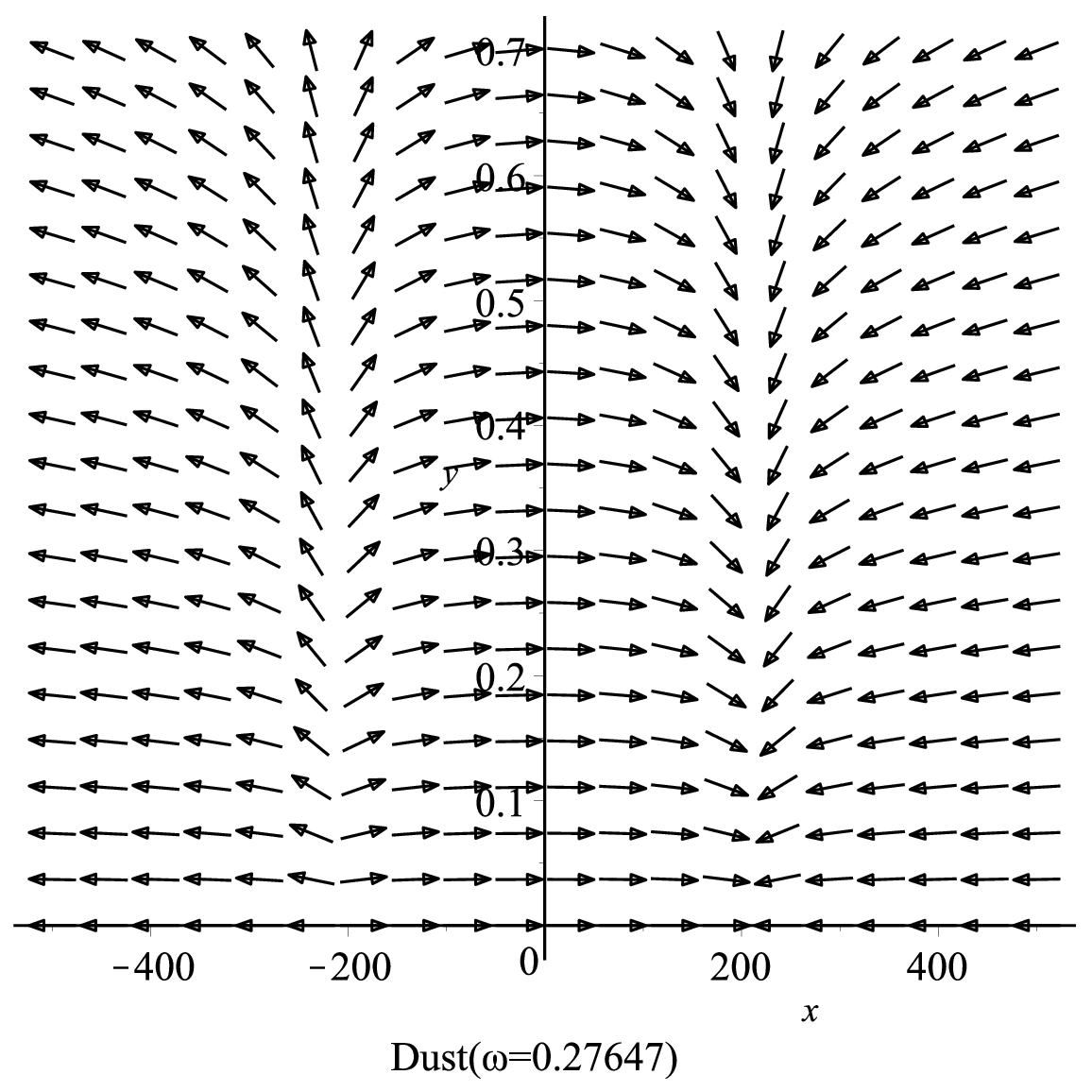}
\includegraphics[width=1.6in,height=1.6in]{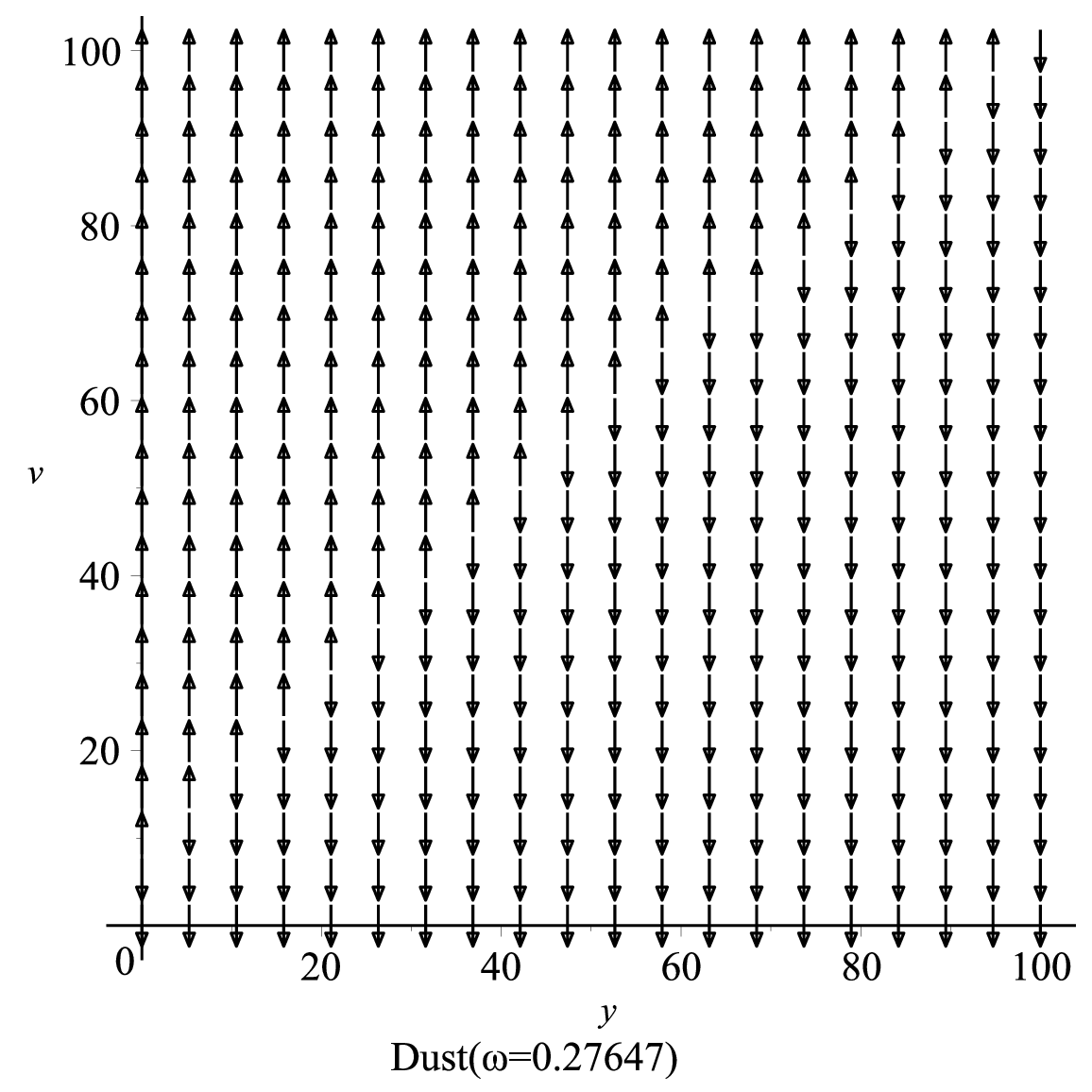}
\includegraphics[width=1.6in,height=1.6in]{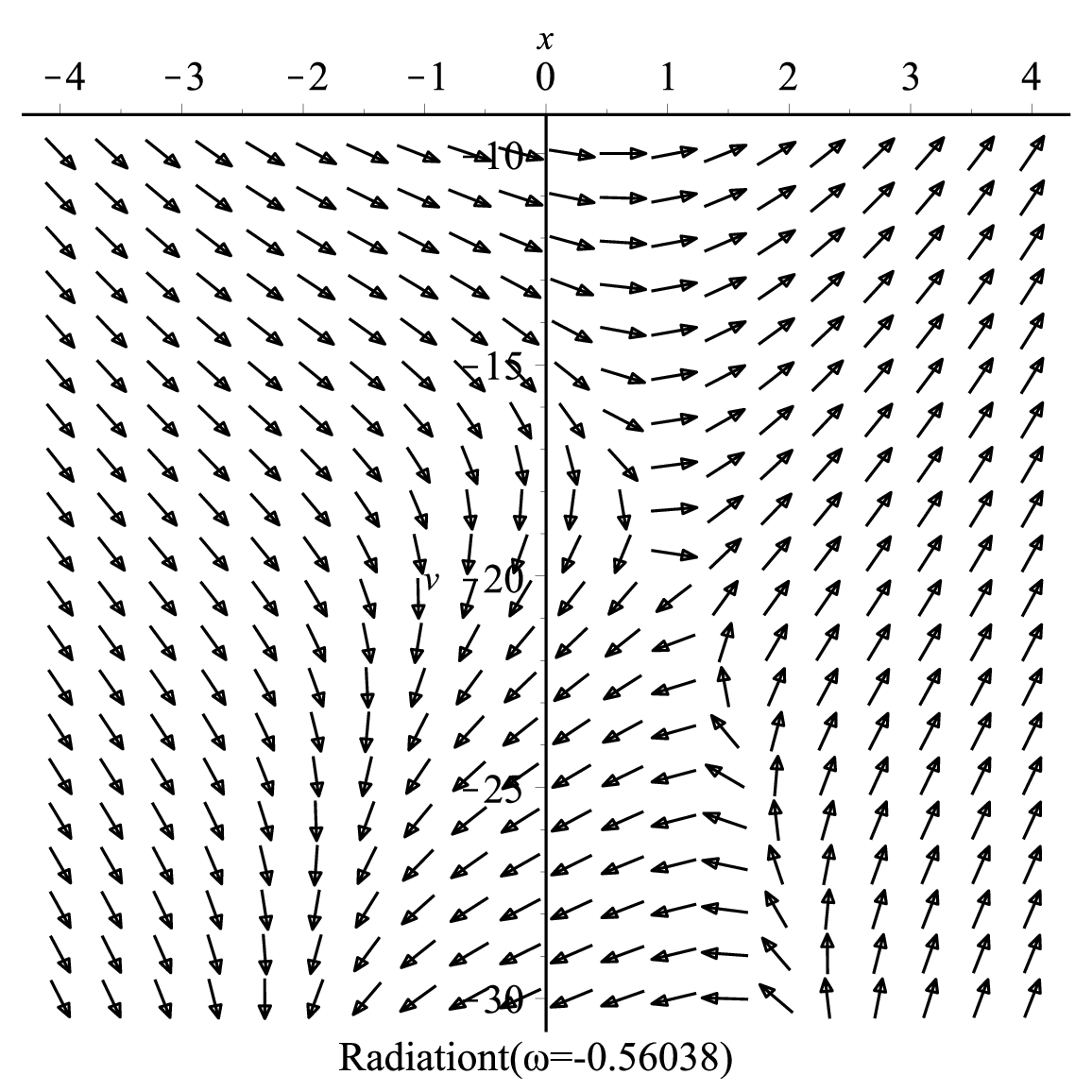}
\includegraphics[width=1.6in,height=1.6in]{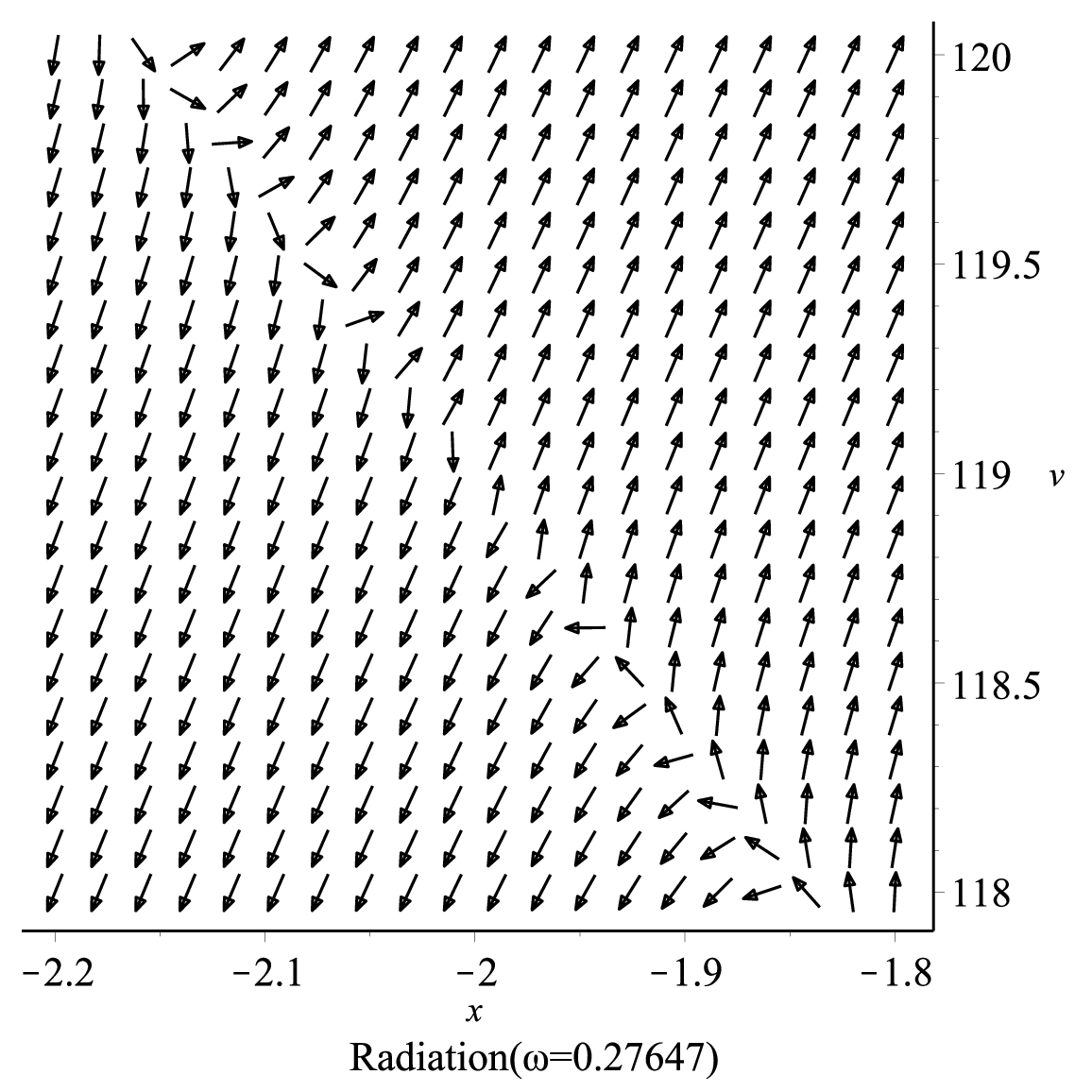}
\includegraphics[width=1.6in,height=1.6in]{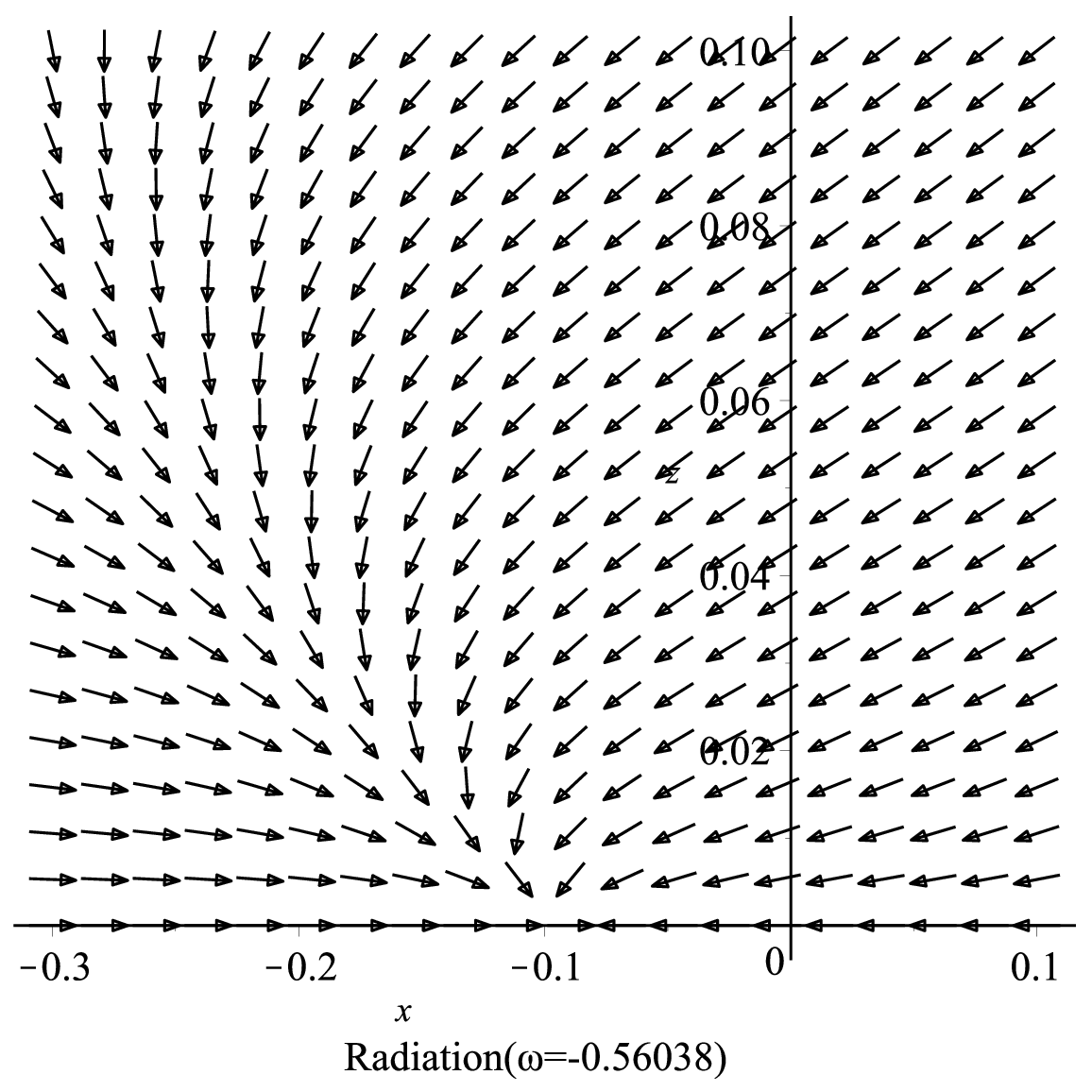}
\caption{{\small \  Arrow diagrams of critical fixed points for de
Sitter, dust and radiation eras }}
\end{figure}

\end{document}